\documentclass[twocolumn,aps,showpacs,superscriptaddress,prl]{revtex4}
\usepackage{hyperref}
\usepackage[dvips]{graphicx}

\begin{document}
\title{Universal patterns in sound amplitudes of songs and music genres}

\author{R.S. Mendes}
\author{H.V. Ribeiro}\email{hvr@dfi.uem.br}
\affiliation{Departamento de F\'isica, Universidade Estadual de
Maring\'a, Av. Colombo 5790, 87020-900, Maring\'a, PR, Brazil}
\affiliation{National Institute of Science and Technology for
Complex Systems, CNPq, Rua Xavier Sigaud 150, 22290-180, Rio de
Janeiro, RJ, Brazil}
\author{F.C.M. Freire}
\affiliation{Departamento de F\'isica, Universidade Estadual de
Maring\'a, Av. Colombo 5790, 87020-900, Maring\'a, PR, Brazil}
\author{A.A. Tateishi}
\author{E.K. Lenzi}
\affiliation{Departamento de F\'isica, Universidade Estadual de
Maring\'a, Av. Colombo 5790, 87020-900, Maring\'a, PR, Brazil}
\affiliation{National Institute of Science and Technology for
Complex Systems, CNPq, Rua Xavier Sigaud 150, 22290-180, Rio de
Janeiro, RJ, Brazil}

\date{\today}

\begin{abstract}
We report a statistical analysis over more than eight thousand songs.
Specifically, we investigate the probability distribution of the normalized
sound amplitudes. Our findings seems to suggest a universal form of distribution
which presents a good agreement with a one-parameter stretched Gaussian.
We also argue that this parameter can give information on music
complexity, and consequently it goes towards classifying songs as well
as music genres. {Additionally, we present statistical evidences that
correlation aspects of  the songs are directly related with the non-Gaussian nature of their
sound amplitude distributions.}
\end{abstract}

\pacs{89.90.+n 89.20.-a 43.75.St 05.45.Tp}

\maketitle

In recent years, studies of complex systems have become
widespread among the scientific community, specially in the
statistical physics one\cite{Auyang,Jensen,Barabasi,Boccara,Sornette}.
Many of these investigations deal with data records ordered in
time or space (i.e., time series), trying to extract some
features, patterns or laws that may be present in the systems
studied. This approach has been successfully applied to a
variety of fields, from physics and
astronomy\cite{Chandrasekhar} to genetics\cite{Peng} and
economy\cite{Mantegna}. Moreover, this framework has been a
trend towards investigating and modeling interdisciplinary
fields, such as religion\cite{Picoli},
elections\cite{Fortunato}, vehicular traffic\cite{Chowdhury},
tournaments\cite{Ribeiro}, and many others. These few examples
and social phenomena in general \cite{Castellano} illustrate as
physicists have gone far from their traditional domain of
investigations.

Music is a well known worldwide social phenomenon linked to the
human cognitive habits, modes of consciousness as well as
historical developments\cite{DeNora}. In the direction of
music's social role, some authors investigated collective
listening habits. For instance, Lambiotte and
Ausloos\cite{Lambiotte} analyzed data from people music library
finding audience groups with the size distribution following a
power law. They also investigated correlations among these
music groups, reporting non-trivial relations\cite{Lambiotte2}.
In another work, Silva et al.\cite{Silva} studied the network
structure of the song writers and the singers of Brazilian
popular music (mpb). There is also an interest in the behavior of music
sales\cite{Lambiotte3} as well as in the success of
musicians\cite{Davies,Borges,Hu}.

Despite these cultural aspects, songs form a highly
organized system presenting very complex structures and
long-range correlations. All these features have attracted the
attention of statistical physicists. In a seminal paper,
Voss and Clarke\cite{Voss} analyzed the power spectrum of radio
stations and observed a $1/f$ noise like pattern. They also
showed that the correlation can extend to longer or shorter
time scales, depending on the music genre. Hs\"u and
Hs\"u\cite{Hsu} investigated the changes of acoustic frequency
in Bach's and Mozart's compositions, finding self-similarity
and fractals structures. In contrast, they report no
resemblance to fractal geometry\cite{Hsu2} for modern music.
Fractal structures have also been reported in the study of
sequences of music notes\cite{Su}, where Su and Wu\cite{Su2}
suggest that the multifractal spectrum can be used to
distinguish different styles of music. By using sound amplitudes of songs, 
Bigerelle and Iost\cite{Bigerelle} achieved a
classification based on fractal dimension using the entire
frequency range. However, as raised by Ro and Kwon\cite{Ro},
the $1/f$ analysis in the region below 20 Hz might not classify
music genres. G\"und\"uz and G\"und\"uz\cite{Gunduz} reported
analysis of several Turkish songs by using many techniques.
Beltr\'an del R\'io et al.\cite{Rio} evaluated the rank
distribution of music notes of a large selection finding a good
agreement with a two parameter beta distribution. Dagdug et
al.\cite{Dagdug} investigated a specific piece of Mozart
employing detrended fluctuation analysis (DFA)\cite{Peng2}.
Applying DFA in a volatility-like series, Jennings et
al.\cite{Jennings} found quantitative differences in the Hurst
exponent depending on the music genre.

\begin{figure}[!t]
\centering
\vspace{-1.3cm}
\includegraphics[scale=0.38]{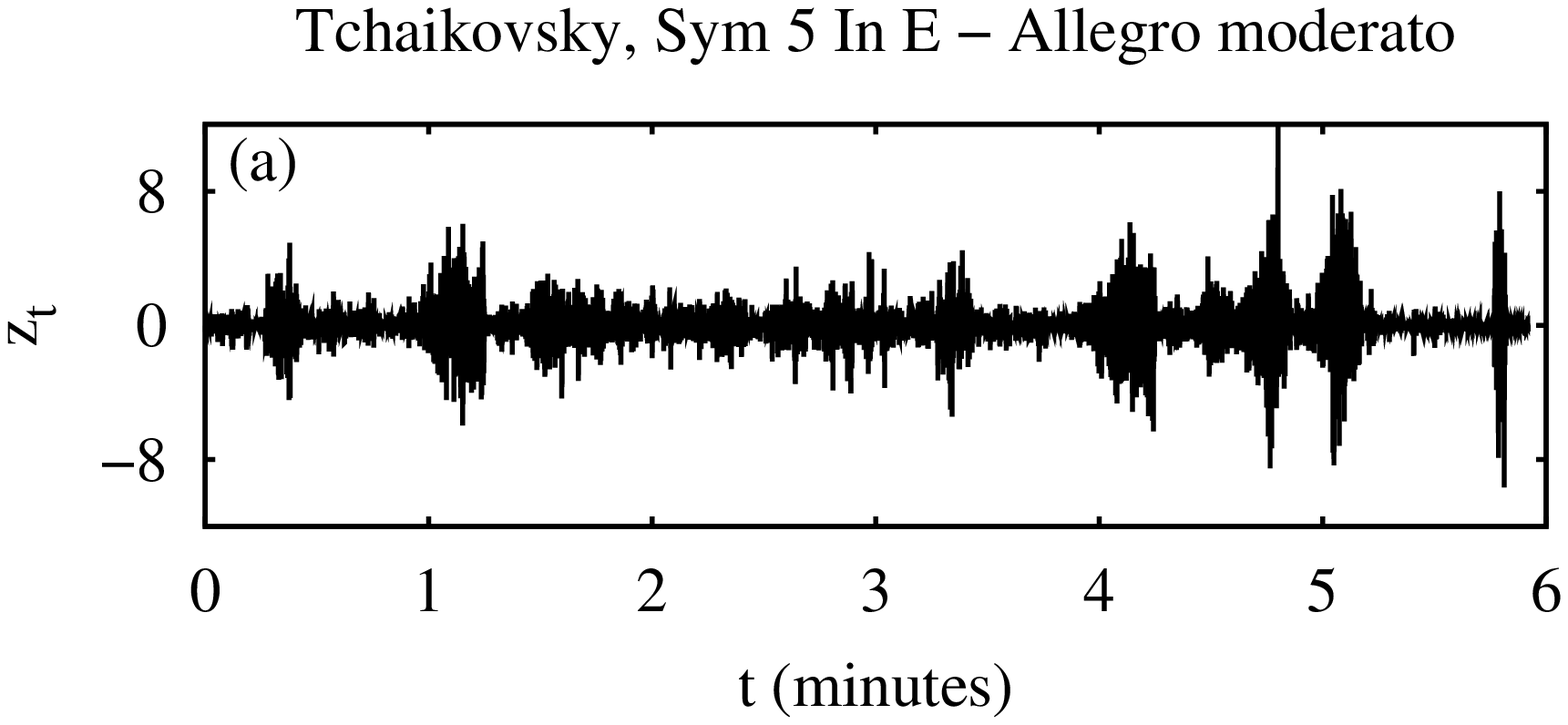}\vspace{-2cm}
\includegraphics[scale=0.38]{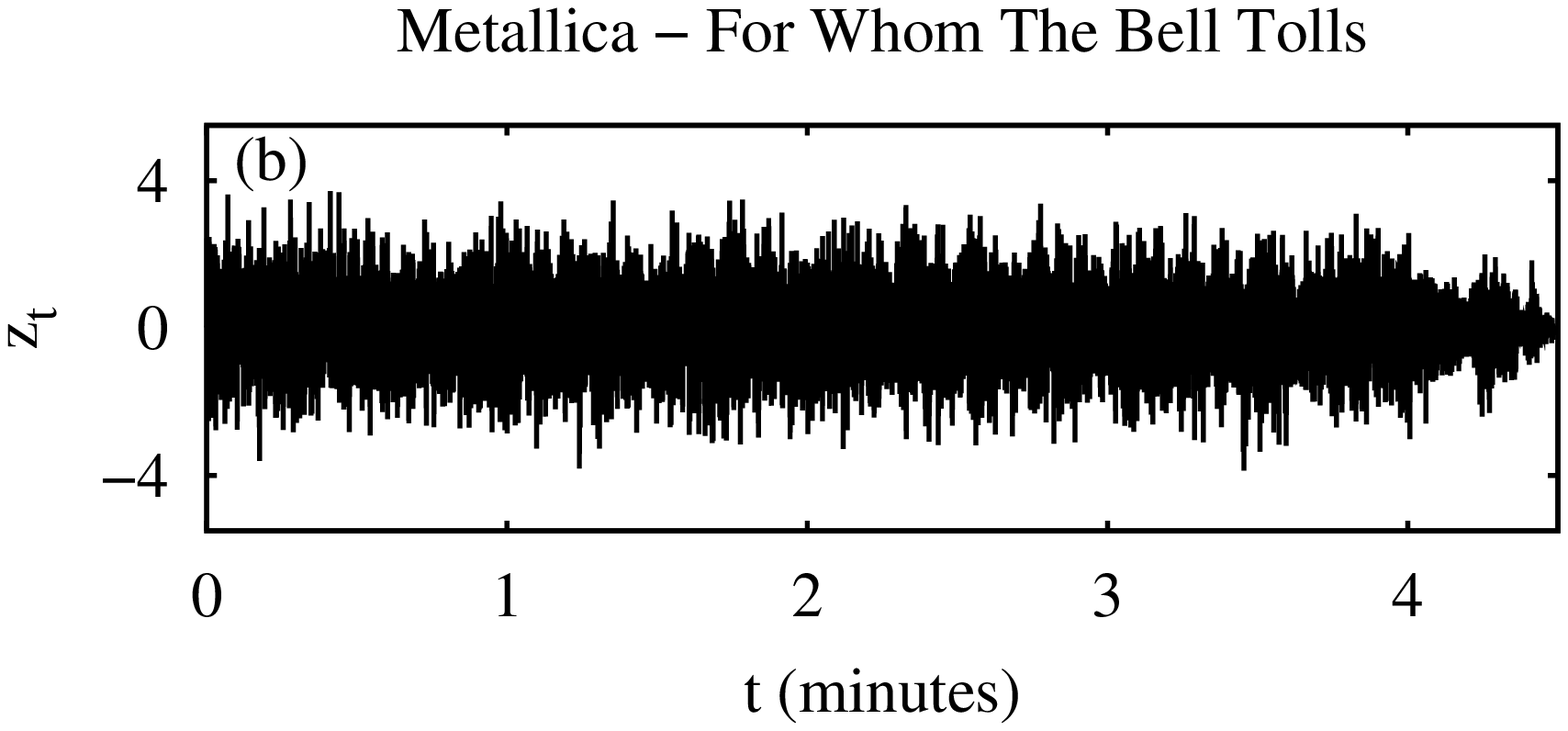}\vspace{-1.1cm}
\caption{The normalized sound amplitude of (a) a classical piece and (b) a heavy metal song (labeled in the figure).
Note that the signals are quite different, the first one presents a more complex structure characterized by ``bursts''
while the second resembles a Gaussian noise.}\label{fig:series}
\end{figure}

In this brief literature review, we see that special attention
was paid to the fractal structures of music, correlations and
power spectrum analysis. However, much less attention has been
paid to the understanding of the amplitude distribution. This last
point has been noted by Diodati and Piazza\cite{Diodati}.
In their work, they investigated the distribution of times and
sound amplitudes larger than a fixed value. By using
this kind of return interval analysis\cite{Santhanam}, they
found Gaussian distributions in the amplitude for jazz, pop,
and rock music, while non-Gaussians emerge for classical
pieces. Here, we directly investigate the amplitude
distributions of songs of several genres without
employing a threshold value as considered by Diodati and
Piazza. Moreover, our analysis goes towards finding patterns
in the amplitude sound distribution by using a suitable one-parameter probability
distribution function (pdf). In the following, we present the dataset 
used in our investigation, the analysis of the shape of the
resulting distributions and our conclusions.

Not all sound is music, but certainly music is made by sounds. The
sounds that we hear are consequence of pressure fluctuations
traveling in the air and hitting our ears. 
These audible pressure fluctuations
can be converted into a voltage signal $u_t$ by using
a record system and stored, for instance, in a compact disk
(CD). Our analysis is focused on this time series $u_t$ that we
call sound amplitude.  In the case of songs stored in
CDs, $u_t$ has a standard sampling rate of 44.1 kHz and
encompasses the full audible human range (approximately between 20
Hz and 20 kHz).

As database we have 8115 songs of nine different music
genres: classical (907), tango (992), jazz (700), hip-hop (876), mpb (580),
flamenco (524), pop (998), techno (900) and heavy metal (1638).  The songs were 
chosen so as to cover a large amount of composers/singers. For instance, for classical
music, we have taken pieces from Bart\'ok, Beethoven, Berlioz, Brahms, Bruch, Chopin, 
Dvorak, Faur\'e, Grieg, Malher, Marcello, Mozart, Rachmaninov, Strauss, 
Schuber, Schumann, Scriabin, Shostakovich, Sibelius, Stravinsky, Tchaikovsky, Verdi, 
Vivaldi, and others.

\begin{figure*}[!t]
\centering
\includegraphics[scale=0.30]{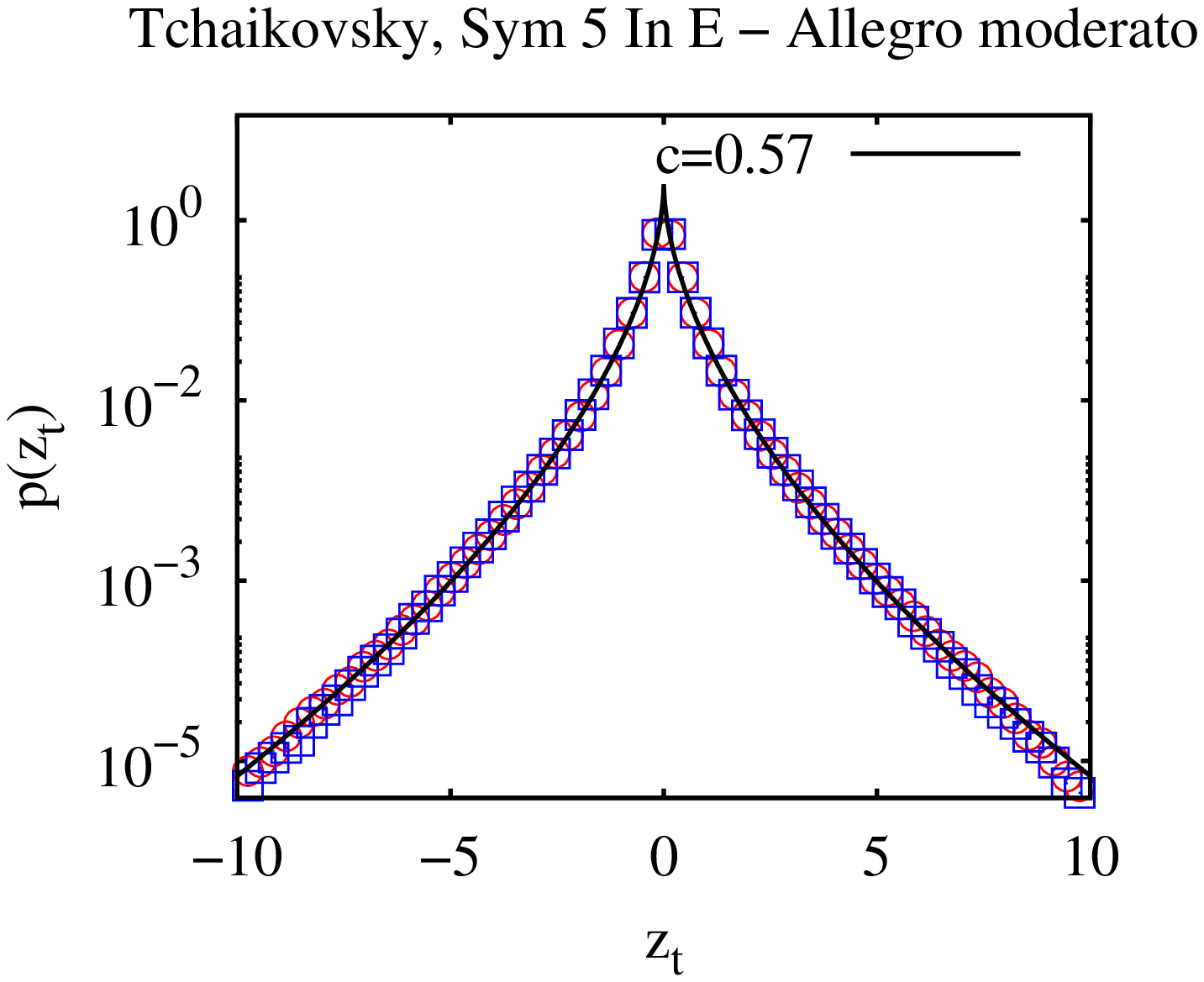}\hspace{-1.7cm}
\includegraphics[scale=0.30]{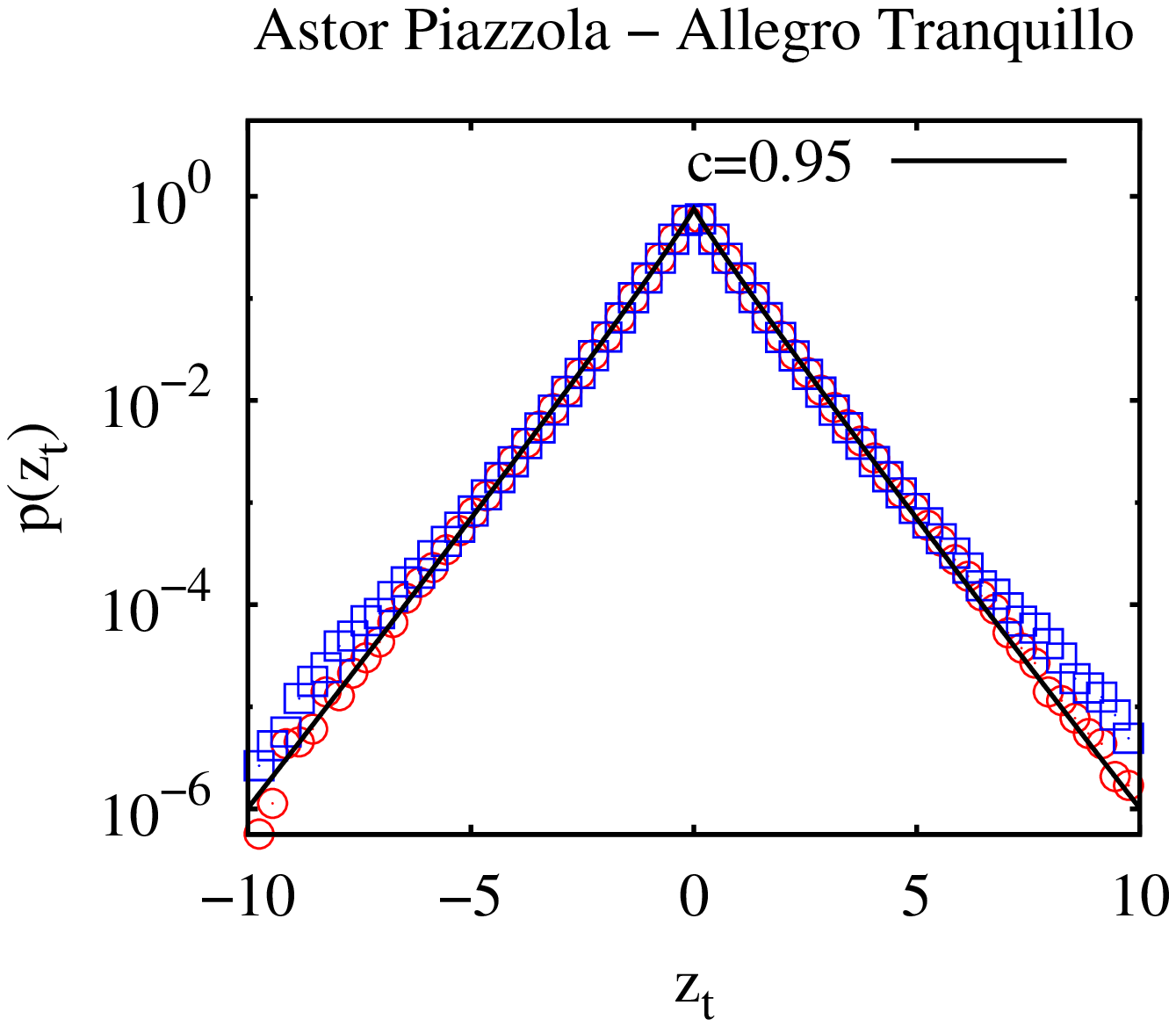}\hspace{-1.7cm}
\includegraphics[scale=0.30]{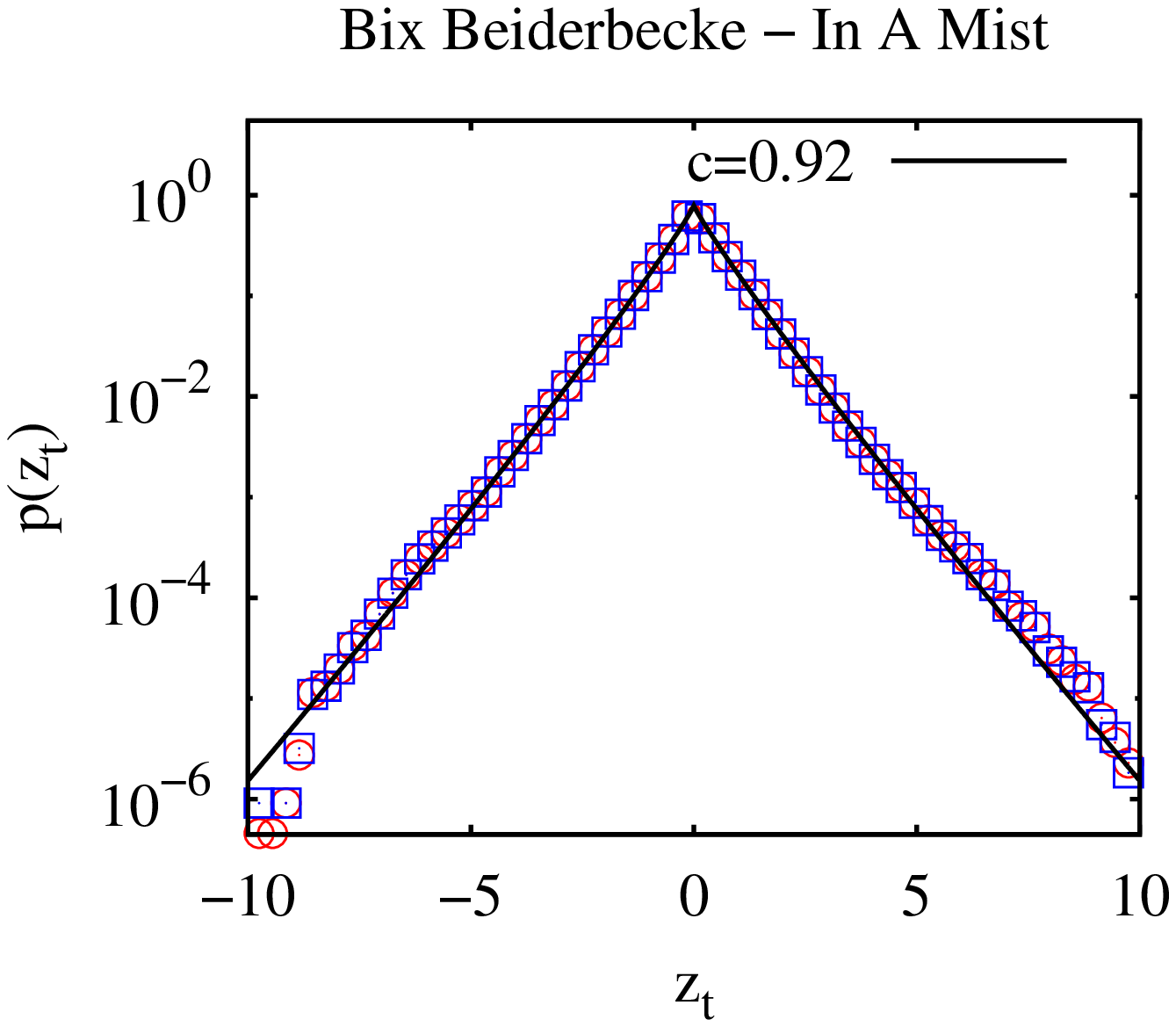}\vspace{-0.3cm}
\includegraphics[scale=0.30]{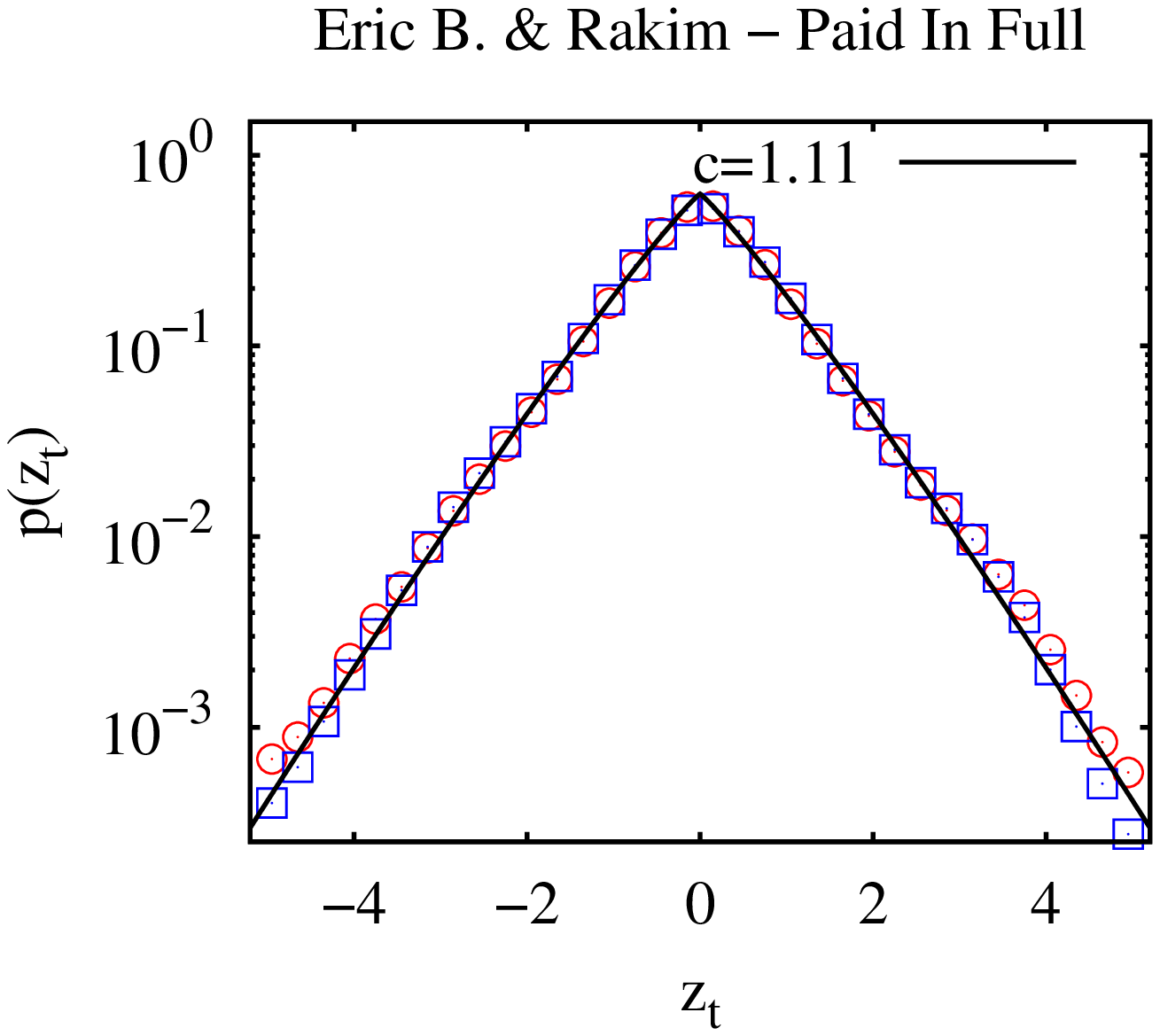}\hspace{-1.7cm}
\includegraphics[scale=0.30]{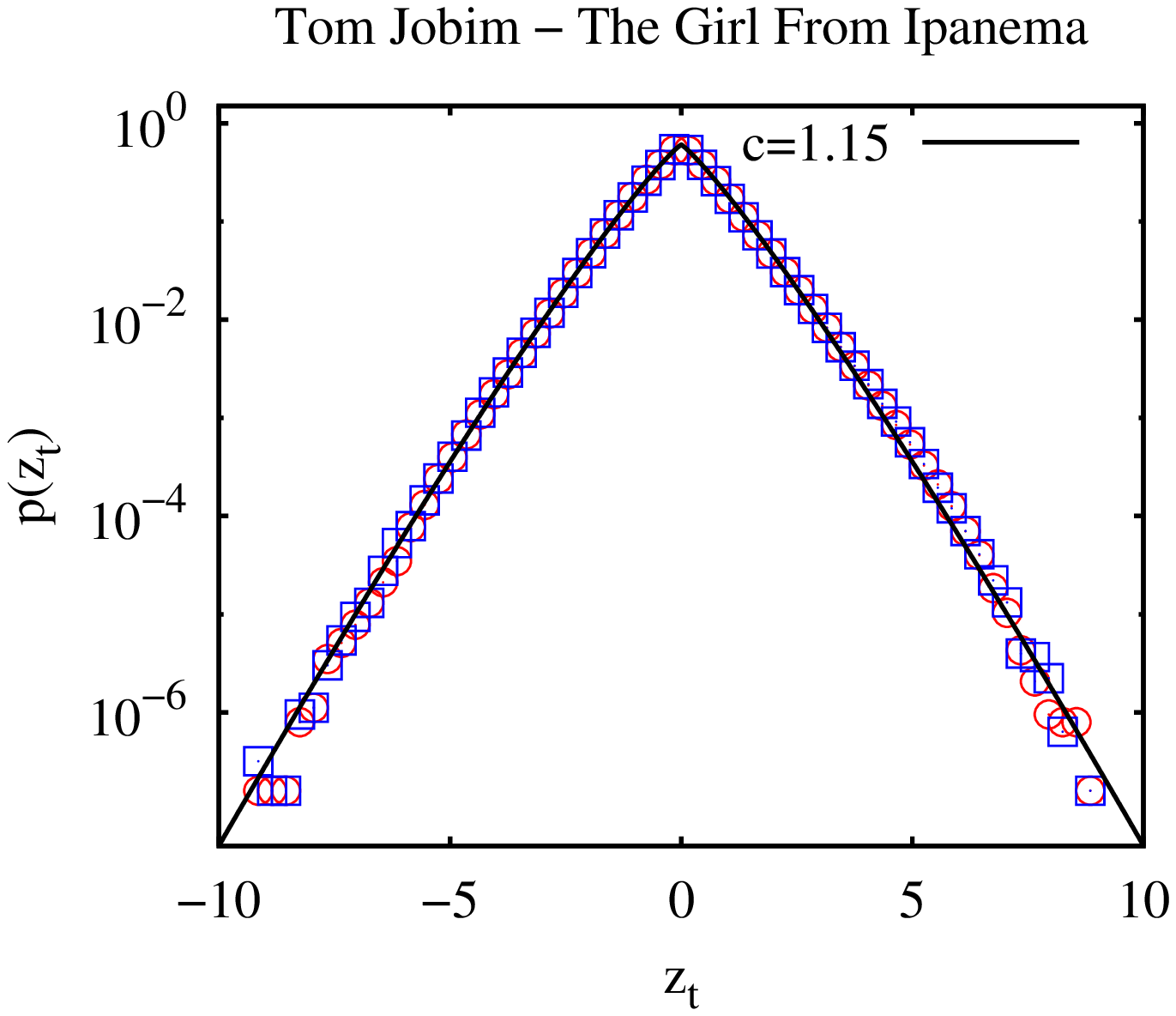}\hspace{-1.7cm}
\includegraphics[scale=0.30]{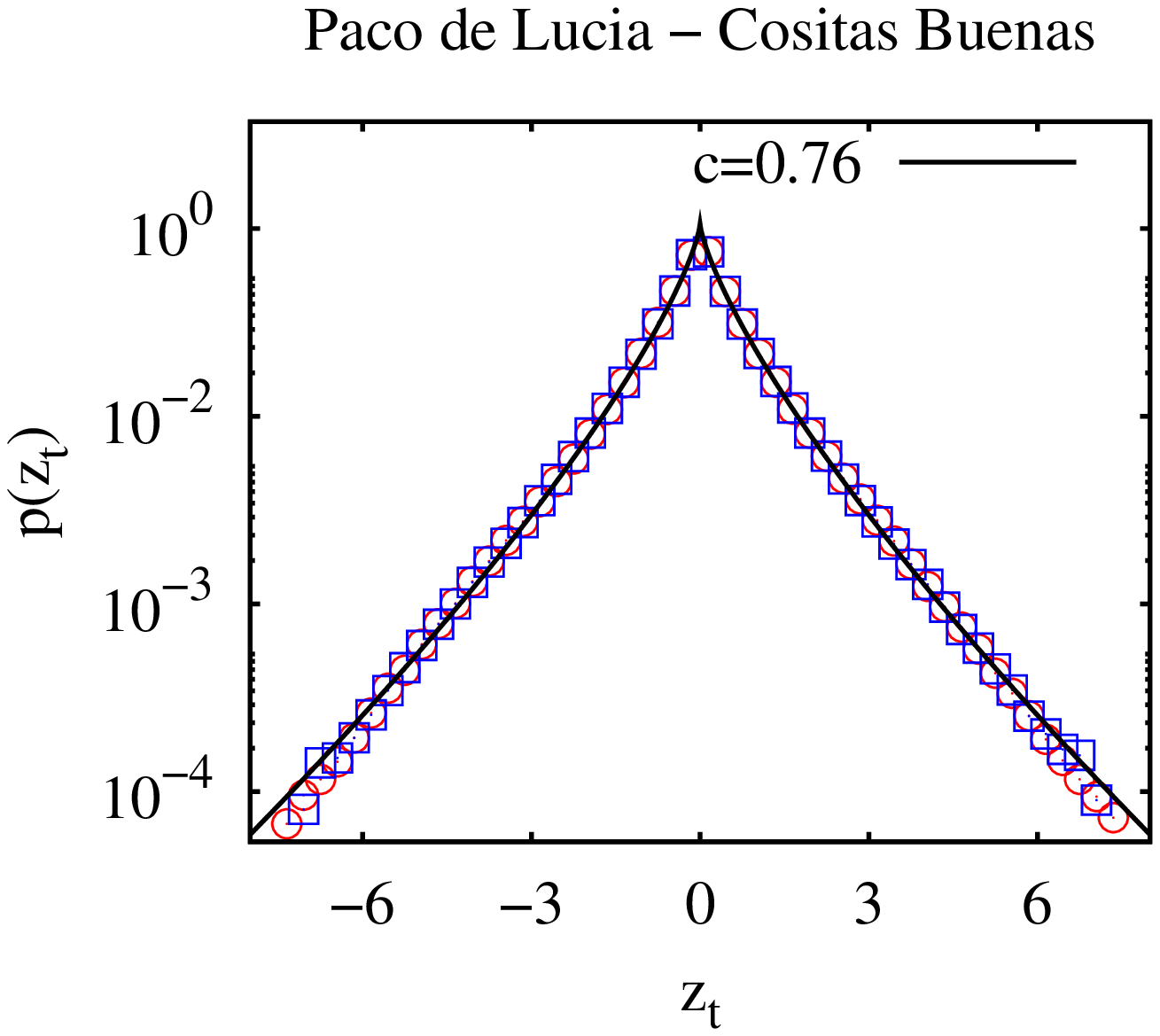}\vspace{-0.3cm}
\includegraphics[scale=0.30]{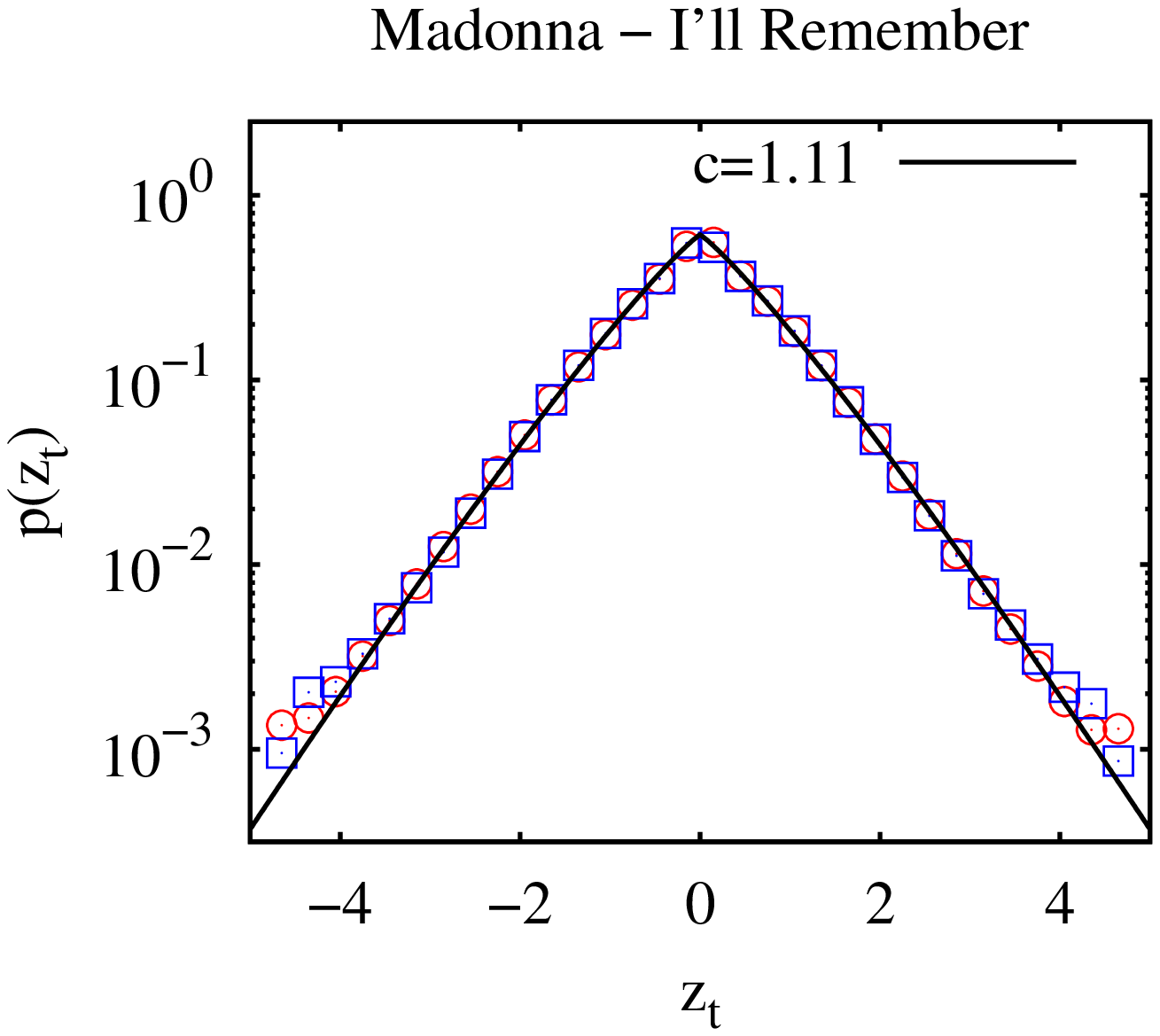}\hspace{-1.7cm}
\includegraphics[scale=0.30]{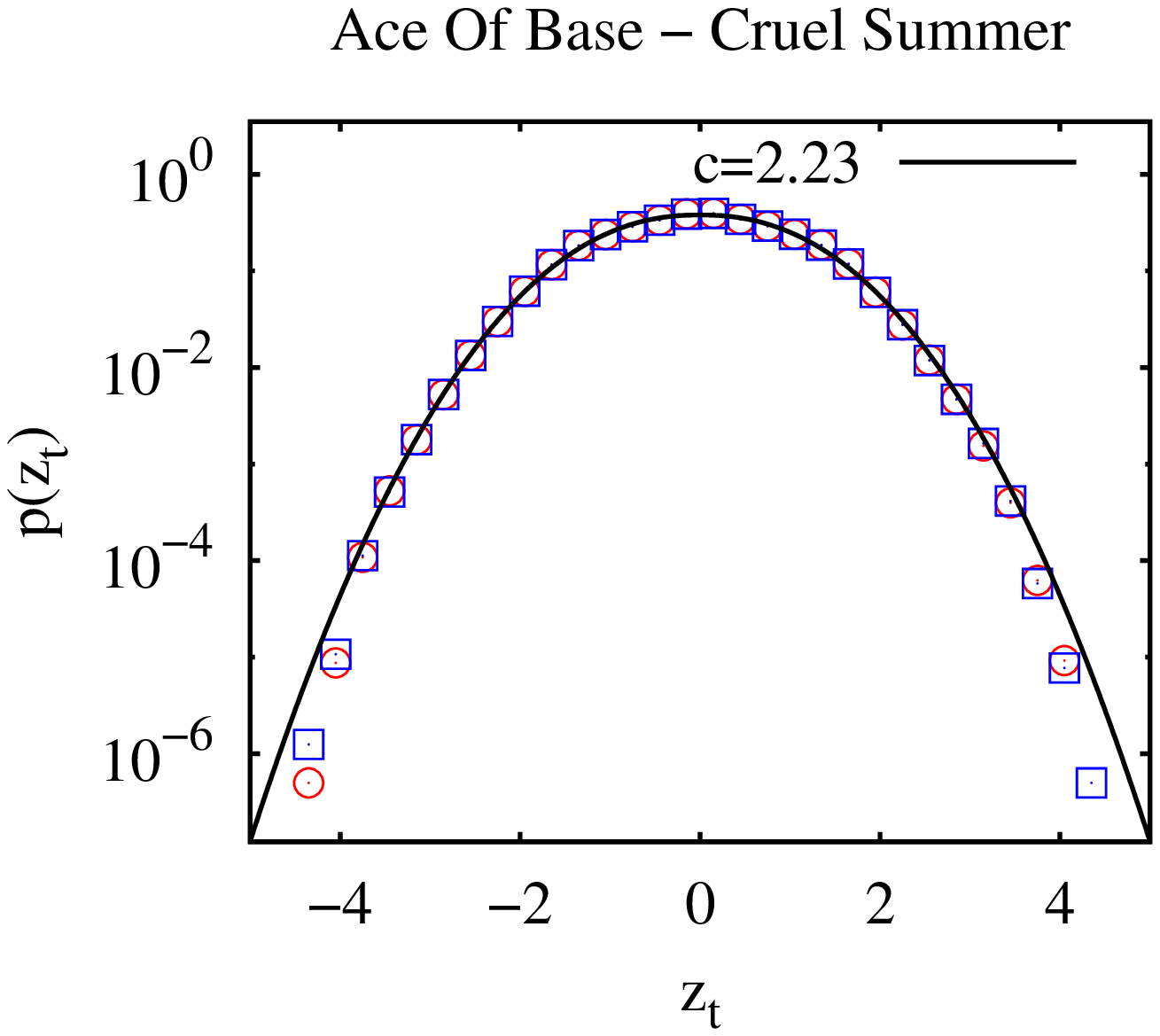}\hspace{-1.7cm}
\includegraphics[scale=0.30]{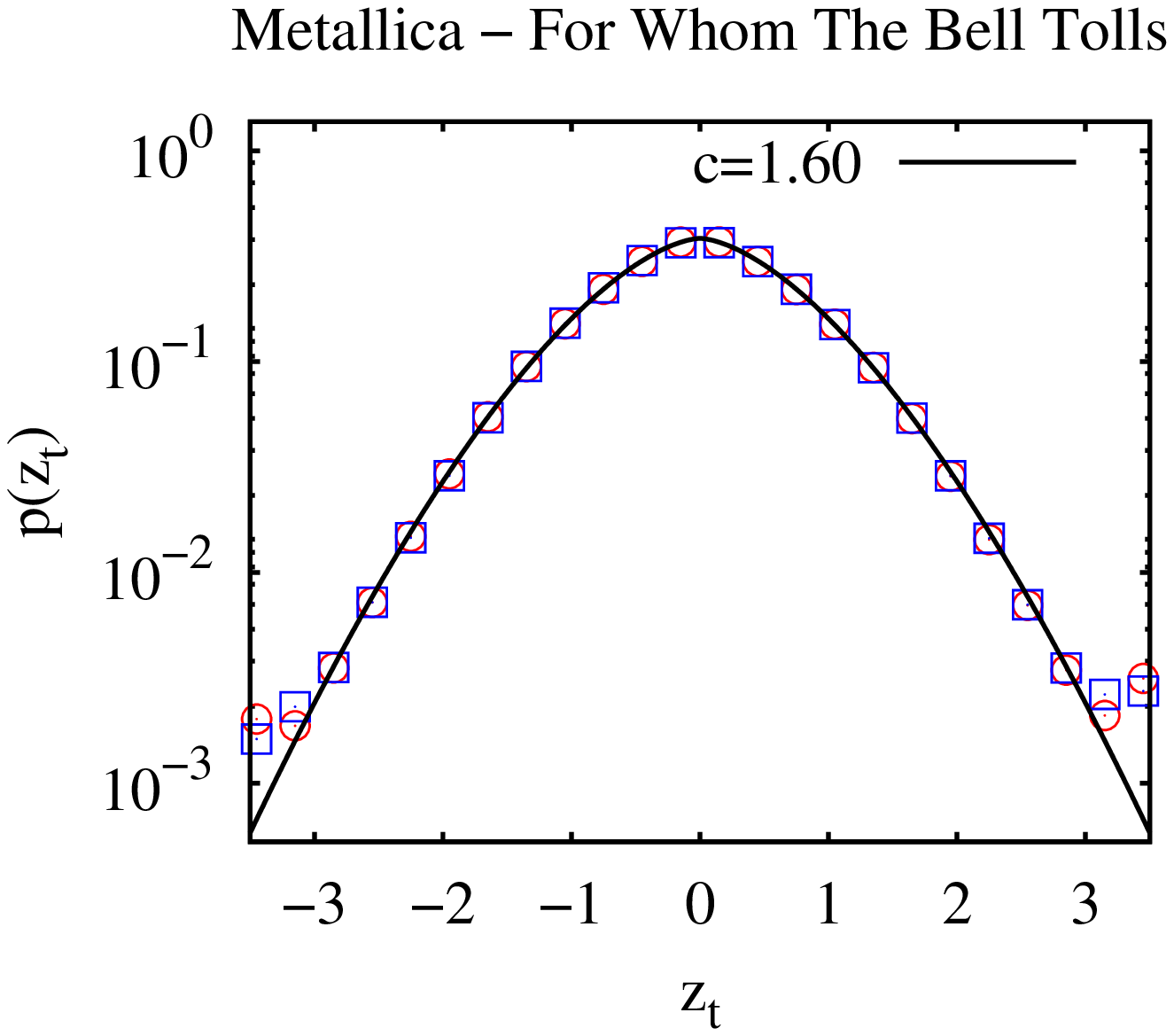}
\caption{(color online) Histograms of some representative songs (labeled in the figure) in comparison with the stretched Gaussian Eq.(\ref{eq2}). The
squares (circles) is the right (left) channel of the stereo audio. As we see, the two channels are quite similar in the sense that the statistical results 
do not dramatically change when considering the right or left channel.}
\label{fig:hist}
\end{figure*}
\begin{figure*}[!ht]
\centering
\includegraphics[scale=0.3]{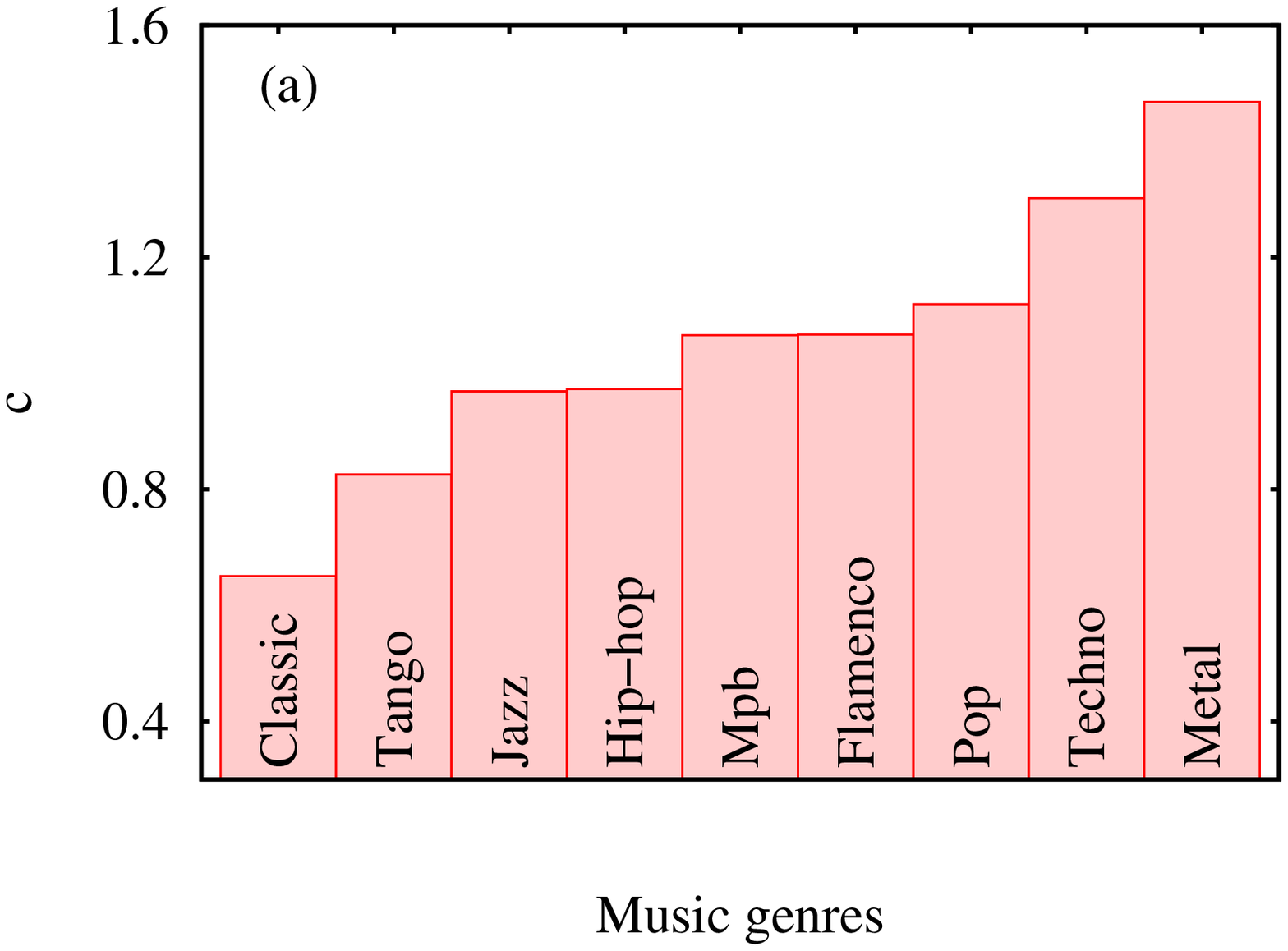}
\includegraphics[scale=0.3]{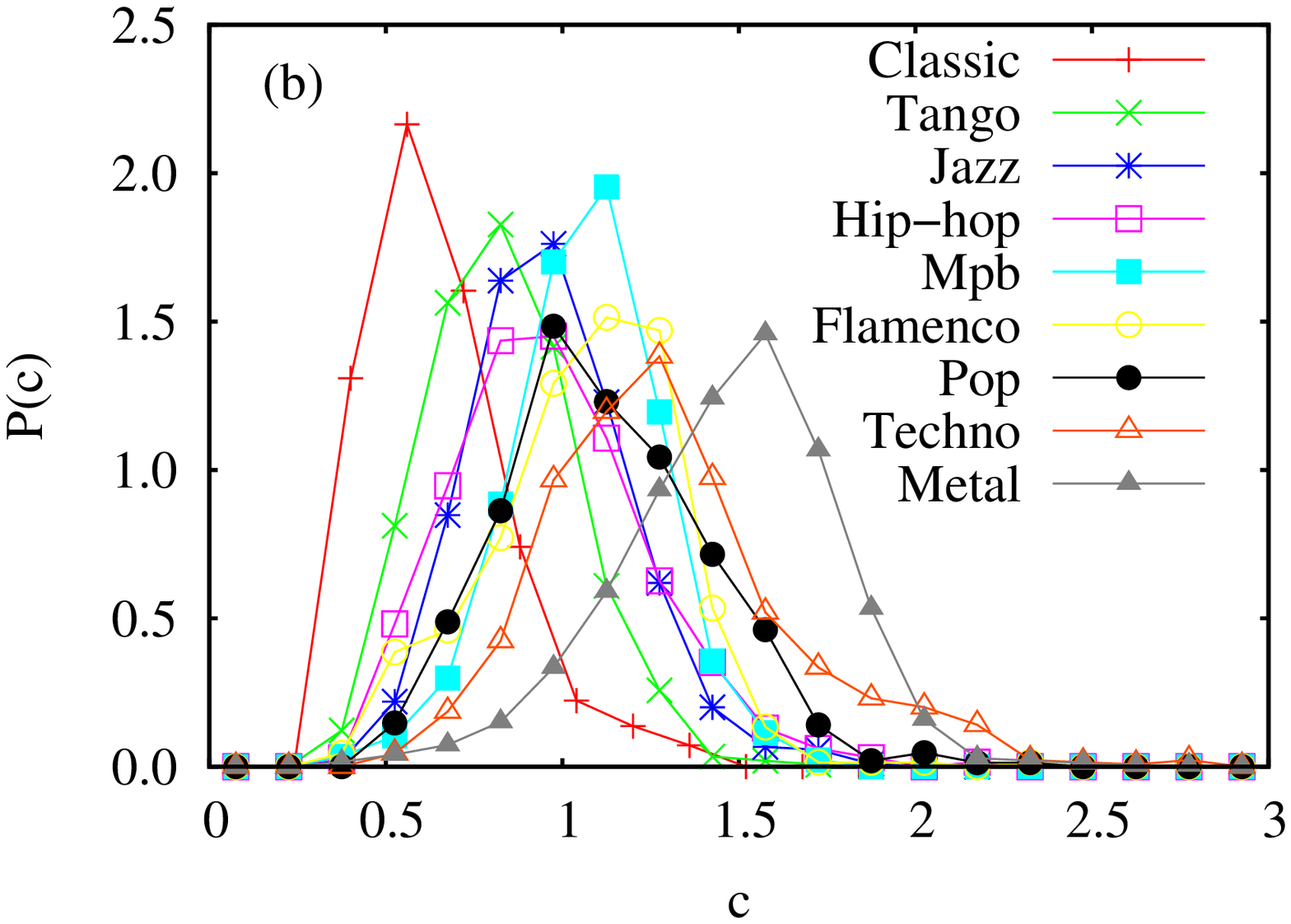}
\includegraphics[scale=0.3]{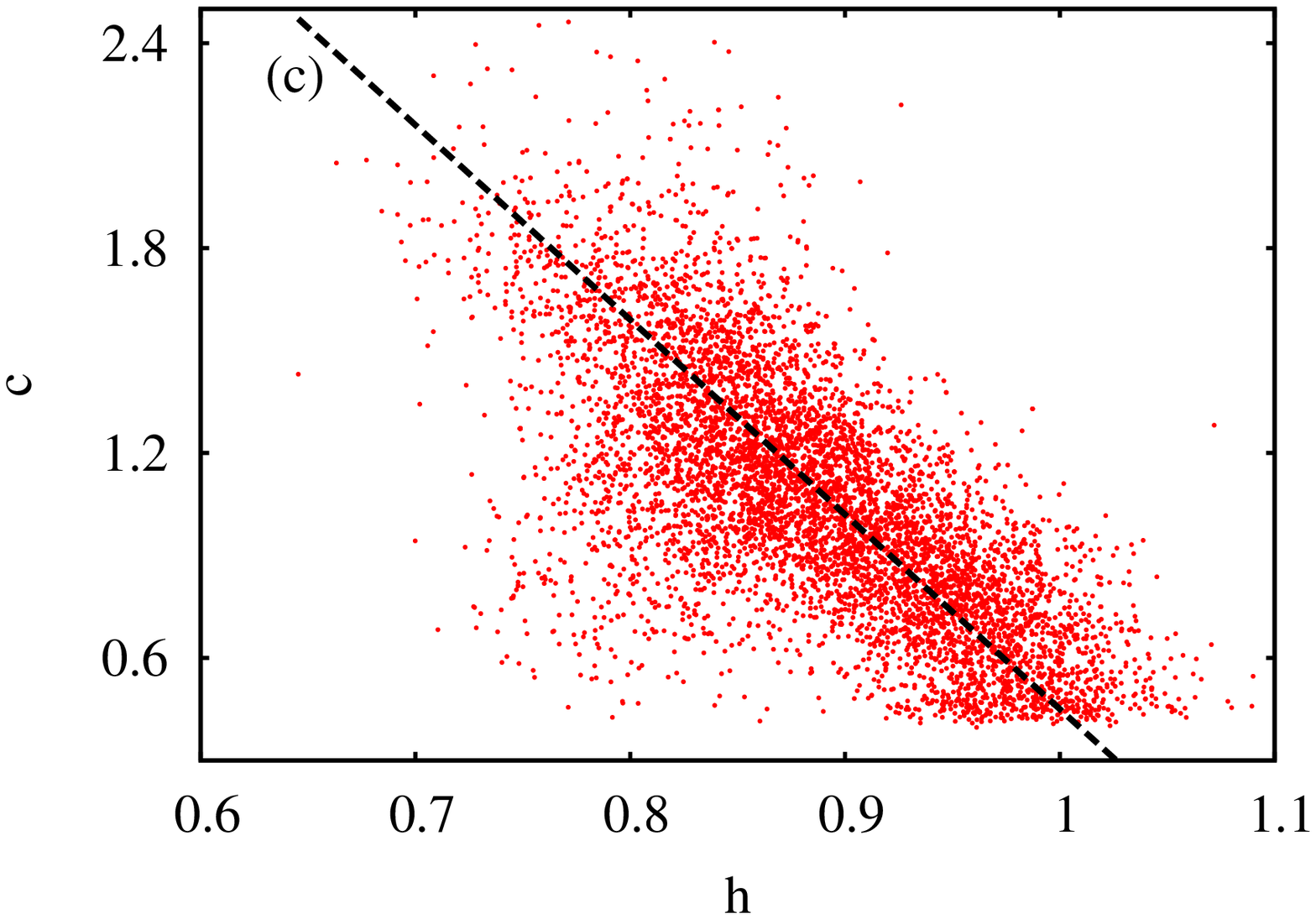}
\caption{(color online) (a) In ascending order, the mean value of the parameter $c$ corresponding
to the stretched Gaussians employed for each music genre considered here. (b) The distribution of the parameter $c$ for each
genre. {(c) Scatter plot of the parameter $c$ versus the Hurst exponent, $h$, obtained via detrended fluctuation analysis (DFA)\cite{Peng3, Kantelhardt} of sound intensity
$z_t^2$. The dashed line is a guide for our eyes.} 
 }
\label{fig:params}
\end{figure*}


When a time series is analyzed, a way to view its variability
(complexity) is at least in part by investigating its pdf. In
the case of music, the mean amplitude is approximately zero
since a vibration essentially occurs around this value. In
addition, the mean (global) intensity is not relevant to the
variability (complexity) of a song. Motivated by these facts,
our research is based on the pdf of recorded data
regardless of their mean value and their real amplitudes. In
other words, we are considering that the complexity of a song
is not related to its mean intensity but with the relative
variability of the amplitudes. Thus, instead of employing the
amplitude $u_t$ in different time instants $t$, we focus
attention on $u_t$ subtracted from its mean value $\mu$ and divided
by its standard deviation $\sigma$. This corresponds to using
$z_t =(u_t - \mu)/\sigma$ instead of $u_t$. Figure \ref{fig:series}
illustrates the behavior of $z_t$ for two songs, a classical piece and a heavy metal song. 
This figure is enough to reveal
qualitative differences between these two songs. In the classical piece, 
we can observe some kind of bursts giving rise to a non-Gaussian
distribution. However, for the heavy metal song,  the signal is very similar 
to a Gaussian noise -- no complex structure is perceptible.

Motivated by these distinct behaviors, we investigate the distribution
of $z_t$ for all the songs in our data set. In Figure \ref{fig:hist},
we show the pdf for some representative songs. As we can verify from this figure, 
the shape of distributions goes from a long tail to Laplace to Gaussian
distribution. A family of functions that has the Gaussian and the Laplace distributions as
particular case is given
by the stretched Gaussian\cite{richardson} $p(z)= N \exp(-b
|z|^c)$, where $N$ is the normalization constant, $b$ is
directly related to the standard deviation and $c$ is a
positive parameter. Since the distribution $p(z)$ is
normalized to unity and the variable $z$ is defined in such way
that its standard deviation is equal to one, the parameters $N$
and $b$ become a function exclusively of $c$, leading to
\begin{equation}
\label{eq2}
 p(z)= \frac{c}{2} \left(\! \frac{ \Gamma (3/c)}{\Gamma (1/c)^{3}} \right)^{\! 1/2}\!\!
~\exp \! \left(\! - \left( \frac{ \Gamma (3/c)}{\Gamma (1/c)} \right)^{\! c/2} \! |z|^c \right) ,
\end{equation}
with $\Gamma[w]$ being the Euler gamma function. Also in Figure \ref{fig:hist} the
least square fits to the data of the above function are shown. Observe that we find a good agreement
between the data and the model for the songs represented in this figure, and a similar
agreement have been found for the others (at least in the central part of the distribution).

The only model parameter is $c$ and it may give useful information about music
complexity. First note that for values of $c$ smaller than one heavy tail distribution
emerge. In some sense, these heavy tails reflect the complex structures that
we see in Figure \ref{fig:series}a, i.e., larger fluctuations. The increasing of $c$
makes the tails shorter and recover some known distributions (Laplace for $c=1$ and 
Gaussian for $c=2$). In this context, a shorter tail indicates that larger fluctuations
become rare, leading to music signal very similar to a Gaussian noise (see Figure \ref{fig:series}b).
From the musical point of view, the word complexity may be related to several
aspects of the song or even with music taste. In present context, it should be viewed
a comparative measurement, i.e., a measure of how the empirical distributions differs from
the Gaussian one.

Based on the above discussion, we may use $c$ to sort the songs and music genres in a kind of
complexity order (smaller $c$ is related to a large complexity). In order to construct this rank for music genres,
we evaluate the mean value of $c$ over all songs of each music genre considered here
as shown in Figure \ref{fig:params}a.  Our findings agree with other works in the sense 
that there is a quantitative difference between classic and light/dancing music\cite{Jennings,Diodati}.
However, it is interesting to emphasize that music genres are not a well defined concept\cite{Scaringella}.
Thus, any taxonomy may be controversial representing an open problem of automatic classification
like other problems of pattern recognition. To take a glance in this complicated problem we also
evaluate the probability distribution of $c$ for each music genre as shown in Figure \ref{fig:params}b. We can
see that there are overlapping regions for all genres, reflecting the fuzzy boundaries existent in the music genre
definition. 

Despite the complex situation that emerges in the problem of automatic genre classification\cite{Correa,Serra,Mostafa,Silla}, our model
is very simple. From the qualitative point of view, the characteristic of songs and music genres is related
with multidimensional aspects like timbre, melody, harmony, rhythm, among others. Thus, as a minimalist model, the
classification presented here must be viewed as a king of global measure for these qualitative aspects. In addition, we have
to note that correlation aspects are lost when we consider only histogram as presented in Figure \ref{fig:hist}. 
In the same way, information is also lost when someone considers only some correlations. {However, we
remark that the results concern to the genre classification, here obtained only by using the pdf of sound amplitude, are in statistical agreement
with other methods based on correlation analysis. This fact seems to suggest a kind of coupling between the correlation
aspects and the non-Gaussian pdfs. Aiming to highlight this feature, we evaluated the Hurst exponent ($h$) of the time series $z_t^2$
and plot it versus the pdf parameter $c$ in Figure\ref{fig:params}c. The data presented in this figure suggest a
approximated liner relation between $c$ and $h$ (Pearson correlation about -0.7),
providing a statistical evidence that the non-Gaussian nature of the pdfs are directly related to the correlations in songs. 
Therefore, } these two complementary aspects and others compose the multidimensional nature of music 
quantification and classification. 

Summing up, 
 we investigated the probability distribution of the normalized sound amplitudes for more than 
eight thousand musical pieces. The empirical findings seem to suggest a universal form of distribution which
showed to be in good agreement with a stretched  Gaussian. Due to the normalization and the standard deviation fixed as one,
our distribution has only one parameter $c$. We argue that this parameter goes towards quantifying the complexity of songs
as well as music genres. In addition to this universal feature, {we presented empirical evidences that non-Gaussian nature
of sound amplitude pdf are related to the correlation aspects. As an application,} we also hope that the distribution of sound 
amplitudes presented here may have implications for stochastic music compositions.

\acknowledgements
We thank CNPq/CAPES for the financial support and the local university radio for kindly providing the songs.


\begin{thebibliography}{99}
\bibitem{richardson} The streched Gaussian
    was introduced in the context of the anomalous diffusion by
    Richardson\cite{richardson1} by considering a spatial
    dependent diffusion coefficient.
\end{thebibliography}
\end{document}